\documentclass[amsmath,amssymb,nofootinbib]{revtex4}

\usepackage{latexsym}
\usepackage{amssymb}
\usepackage{amsfonts}
\usepackage{amsmath}
\usepackage[dvips]{graphicx}
\usepackage{bm}

\def\rp#1#2{{#1\over#2}}
\def\dert#1#2{\frac{{{d}}{#1}}{{{d}}{#2}}}              
\def\lb#1{\label{#1}}

\def\hb#1{\hat{\mathbf{#1}}}
\def\vb#1{{\mathbf{#1}}}

\def\ber{\begin{eqnarray}}
\def\eer{\end{eqnarray}}
\def\beq{\begin{equation}}
\def\eeq{\end{equation}}

\begin{document}

\title{Perturbations of Keplerian Orbits in Stationary Spherically Symmetric Spacetimes}

\author{Matteo Luca Ruggiero}
\email{matteo.ruggiero@polito.it}
\affiliation{DISAT - Dipartimento di Scienza Applicata e Tecnologia,  Politecnico di Torino, Corso Duca degli Abruzzi 24, Torino, Italy}
\affiliation{INFN, Sezione di Torino, Via Pietro Giuria 1, Torino, Italy}

\date{\today}

\begin{abstract}
We study spherically symmetric perturbations determined by  alternative theories of gravity to the gravitational field of a central mass in General Relativity. In particular, we focus on perturbations in the form of power laws and calculate their effect on the secular variations of the orbital elements of a Keplerian orbit. We show that, to lowest approximation order, only the argument of pericentre and mean anomaly undergo secular variations; furthermore, we calculate the variation of the orbital period. We give analytic expressions for these variations which can be used to constrain the impact of alternative theories of gravity.
\end{abstract}

\maketitle

\section{Introduction} \label{sec:intro}

In recent years, there has been much interest in studying theories of gravity alternative to General Relativity (GR) \cite{tsu,sath,clifton,yoo}: these theories have been motivated by the current observations, which seem to question the GR model of gravitational interactions on large scales, such as the galactic and cosmological ones. Just to mention some examples, one can consider $f(R$) theories of gravity \cite{capozfranc07,Soti,defelice}, MOdified Newtonian Dynamics (MOND) \cite{bek},  $f(T)$ gravity \cite{crc},  braneworld models \cite{dgp}, the relativistic MOdified Gravity theory (MOG) \cite{mog1,mog2}, curvature invariant models \cite{navarro1,navarro2}, Ho\v{r}ava-Lifshitz  gravity \cite{horava1,horava2,horava3}.

In order to get a deeper insight into these theories, it is important to test their predictions in a suitable weak-field and slow-motion limit: to this end, for instance, central mass solutions which generalize the one obtained in GR (the Schwarzschild solution) are investigated and their predictions compared to the data available from astronomical and astrophysical observations, in the Solar System and beyond. For these purposes,  a general theoretical framework that can deal with classes of metric theories of gravity has been developed: the parametrized  post-Newtonian (PPN) formalism. In the latter,  the weak-field and slow-motion  limit of these theories is studied in terms of suitable parameters, in such a way that experiments may fix their values:  as a matter of fact,  the
current best estimates of the PPN parameters are in agreement with GR \cite{Will}. As a consequence,  it is reasonable to believe that the effects of theories alternative to GR are small, so that they can be dealt with as perturbations of the GR background.

In this paper we want to suggest a simple approach that, without requiring the complex and comprehensive PPN framework,  allows to evaluate the predictions of alternative theories of gravity on the Keplerian orbit of a test particle, which can be thought of as a simplified model of the dynamics of celestial bodies in planetary systems. In particular, we consider a general stationary and spherically symmetric (SSS) spacetime metric, that can be thought of a solution of the field equations of a generic alternative theory of gravity describing the gravitational field around a point-like central mass, and we work out the perturbations of the Keplerian orbital elements. We focus on perturbations in form of power laws, which have been recently considered in the literature \cite{adkins,adel1,randall,most,ferrer1,ferrer2,dobre,adel2,iorioA,iorioB}: these perturbations are interesting because they reproduce some known solutions of the field equations of alternative theories of gravity; moreover arbitrary spherically symmetric perturbations can be written in terms of power series, so that our results can be used quite generally. In particular, by using the  Gauss perturbation scheme, we obtain, to lowest approximation order, the expressions of the secular variations of the orbital elements and the orbital period in terms of hypergeometric functions. These expressions can be used to place bounds on the parameters of alternative theories of gravity.

The paper is organized as follows: in Section \ref{sec:geo} we write the perturbing acceleration in a generic SSS spacetime and obtain its expression to lowest approximation order, while the Gauss perturbation equations  are briefly reviewed in Section \ref{sec:pert}. In Section \ref{sec:powerlaw} we focus on perturbations in the form of power laws, while conclusions are in Section \ref{sec:conc}.

\section{The perturbing acceleration} \label{sec:geo}

We suppose that the gravitational field of a point mass in a generic alternative theory of gravity is described by the SSS spacetime metric\footnote{If not otherwise stated, we use units such that $c=G=1$; 
 bold face letters like ${\mathbf{x}}$ refer to spatial vectors while Latin indices refer to spatial components.}
\beq
ds^{2}=\left(1+\phi(r)\right)dt^{2}- \left(1+\psi(r)\right)\left(dr^{2}+r^{2}d\Omega^{2} \right) \label{eq:metrica00}
\eeq
where $d\Omega^{2}=d\theta^{2} +\sin^{2} \theta d\varphi^{2}$, and $\phi(r),\ \psi(r)$ are functions depending on the mass $M$ of the source and, possibly, on other parameters of the theory.

We point out that, since we are considering spherically symmetric spacetime, gravitomagnetic effects \cite{gem1} are intentionally neglected: in other words in calculating the test particle equation of motion (see equation (\ref{eq:geod1}) below), we do not take into account the perturbing acceleration deriving from the off-diagonal element of the spacetime metric (see e.g. \cite{barker,iorioC,renzetti} for gravitomagnetic perturbations of the orbital elements) 

The metric (\ref{eq:metrica00}) is written in isotropic polar coordinates, such that the spatial part of the metric is proportional to the flat spacetime metric $dr^{2}+r^{2}d\Omega^{2}=dx^{2}+dy^{2}+dz^{2}$. Consequently, it is possible to write the metric (\ref{eq:metrica00}) in Cartesian coordinates in the form
\beq
 ds^{2}=\left(1+\phi(r)\right)dt^{2}- \left(1+\psi(r)\right)\left(dx^{2}+dy^{2}+dz^{2}\right) \label{eq:metrica01}
\eeq
where $r=\sqrt{x^{2}+y^{2}+z^{2}}$. 

Since the effects of the generic alternative theory of gravity are expected to be small, we suppose that they can be considered as perturbations of the known GR solution, i.e. the Schwarzschild spacetime. This amounts to saying that $\phi(r)$ and $\psi(r)$ must approach their GR values: in other words we suppose that in a suitable limit, the metric takes the form
\begin{eqnarray}
\phi(r) & = & \phi_{GR}(r)+\phi_{A}(r) \label{eq:defar1} \\
\psi(r) & = & \psi_{GR}(r)+\psi_{A}(r) \label{eq:defbr1}
\end{eqnarray}
where the GR values are given by $\phi_{GR}(r)=-2M/r+2M^{2}/r^{2}$, $\psi_{GR}(r)=2M/r$ (see e.g. \cite{sereno03}) and the perturbations $\phi_{A}(r), \psi_{A}(r)$ due to the alternative gravity model are such that $\phi_{A}(r) \ll \phi_{GR}(r)$, $\psi_{A}(r) \ll \psi_{GR}(r)$.

In order to calculate the perturbations of the orbital elements, we must calculate the perturbing acceleration. To this end, first we assume that in the given theory, the matter is minimally and universally coupled, so that test particles follow geodesics of the metric (\ref{eq:metrica01}) (or, equivalently (\ref{eq:metrica00})).  Then, we consider the (post-Newtonian) equation of motion of a test particle (see \cite{brumberg})
\beq
\ddot x^{i}=-\frac 1 2 h_{00,i}- \frac 1 2 h_{ik}h_{00,k}+h_{00,k}\dot x^{k} \dot x^{i}+ \left(h_{ik,m}-\frac 1 2 h_{km,i} \right)\dot x^{k} \dot x ^{m} \label{eq:geod1}
\eeq
where ``dot'' stands for derivative with respect to the coordinate time. Since  in our notation  it is $h_{00}=\phi(r)$ and $h_{ij}=\psi(r)\delta_{ij}$, we can write the perturbing acceleration $\vb{W}$   in the form
\beq
\vb{W}=-\frac 1 2 \left \{\Phi(r) \left[1+\psi_{A}(r) \right]+\Psi(r)v^{2}  \right \} \hb x+ \left[\Phi(r)+\Psi(r) \right]\left(\hb x \cdot \vb v \right) \vb v \label{eq:acc1}
\eeq
where we set $\vb x= \left(x,y,z \right)$, $\vb v=\left(\dot x, \dot y, \dot z \right)$, $\hb x= \vb x/|\vb x|$, and
\beq
\Phi(r) \doteq \frac{d \phi_{A}(r)}{dr}, \quad \quad \Psi(r) \doteq \frac{d \psi_{A}(r)}{dr} \label{eq:defAB1}
\eeq

We aim at investigating the lowest order effects on planetary motion of $\phi_{A}(r)$ and $\psi_{A}(r)$: to this end, it is sufficient to apply the  Gauss perturbation scheme to a Keplerian ellipse. 

Moreover, since we are interested in the lowest order effects, we may also neglect the non linear terms (i.e. non linear perturbations with respect to flat spacetime). To this end, we start by noticing that  to Newtonian order, it is $v^{2} \simeq \phi_{GR}(r) \simeq M/r$. As a consequence,  in (\ref{eq:acc1}) we may neglect  the terms proportional to $v^{2}$ and to $\left(\hb x \cdot \vb v \right) \vb v$ (which is also proportional to the orbital eccentricity) and also the term $\Phi(r) \psi_{A}(r)$. In summary, in the weak-field and slow-motion limit the perturbing acceleration that we are going to consider is purely radial and is  given by 
\beq
\vb{W}=-\frac{1}{2} \Phi(r) \hb x  \doteq W_{r} \hb x \label{eq:acc3}
\eeq

We point out that even though we started from the study of a SSS spacetime with the aim of setting bounds on alternative theories of gravity, what follows can be applied to generic radial perturbations (in particular, to those in form of  a power law) of  the orbital elements of a Keplerian motion.

\section{Perturbation Equations} \label{sec:pert}

We start from the expression (\ref{eq:acc3}) of the acceleration and  apply the  Gauss perturbation scheme to obtain  the secular variations.  Because of spherical symmetry, the motion of test particles is confined to a plane and, in this plane,   the unperturbed Keplerian ellipse is
\beq
r= \frac{a\left(1-e^{2}\right)}{1+e\cos f} \label{eq:ellipse1}
\eeq
where $a$ is the semimajor axis, $e$ the eccentricity, $f$ the true anomaly describing the particle angular distance from the pericentre.
For purely radial perturbations, the  Gauss equations  for the variations of the Keplerian orbital elements read \cite{bfv}
\begin{eqnarray}\lb{Gauss}
\dert a t & = & \rp{2e}{n\sqrt{1-e^2}}  W_r\sin f ,\lb{gaus_a}\\
\dert e t  & = & \rp{\sqrt{1-e^2}}{na} W_r\sin f, \lb{gaus_e}\\
\dert I t & = & 0,\lb{gaus_I}\\
\dert\Omega t & = & 0, \lb{gaus_O}\\
\dert\omega t & = &-\rp{\sqrt{1-e^2}}{nae} W_r\cos f,\lb{gaus_o}\\
\dert {\mathcal{M}} t & = & n - \rp{2}{na} W_r\left(\rp{r}{a}\right) -\sqrt{1-e^2}\dert\omega t ,\lb{gaus_M}
\end{eqnarray}
where ${\mathcal{M}}$  is mean anomaly, $I$ is the orbital inclination, $\Omega$ is  the ascending node, $\omega$ is the argument of pericentre,   $n =\sqrt{M/a^3}$ is the Keplerian mean motion. In a Keplerian orbit the mean anomaly is a linear function of time  defined by $\mathcal{M}=n\left(t-t_{0}\right)$, where   $t_{0}$ is the time of a passage through pericentre.  The orbital period $P_{b}$  is related to the mean motion by $n=2\pi /P_{b}$.

In order to obtain the secular effects  we must evaluate the Gauss equations onto the unperturbed Keplerian ellipse (\ref{eq:ellipse1}),  and then we must average them over one orbital period of the test particle. From (\ref{gaus_a}) and (\ref{gaus_e}), and taking into account (\ref{eq:ellipse1}), we see that the secular variations of the semimajor axis and eccentricity are null, because when averaging them the arguments of the integrals are odd functions. Hence, we obtain that in SSS spacetimes,  to lowest approximation order (i.e. when the perturbations are purely radial)  semimajor axis, eccentricity, beside the  inclination and the node (which are only affected by normal i.e. out-of the-plane perturbations), are not affected by secular variations. We point out that, once that the secular variations of the argument of the pericentre $<\dot{\omega}>$ and mean anomaly $<\dot{\mathcal{M}}>$ have been determined, it is possible
to obtain the corresponding variation of the mean longitude $<\dot{\lambda}> \doteq <\dot{\omega}> +<\dot{\mathcal{M}}>$ \cite{bfv}.

Starting from the equation describing the variation of the mean anomaly, it is possible to evaluate the perturbation of the orbital period.  In fact (see \cite{iorio07a}), on using 
\beq
df= \left (\frac a r \right)^{2} \sqrt{1-e^{2}}d\mathcal M \label{eq:dfdM1}
\eeq
from (\ref{gaus_M}) it is possible to write
\beq
\dert f t=n  \left (\frac a r \right)^{2} \sqrt{1-e^{2}}  \left[1-\frac{2}{n^{2}a} W_{r} \left(\frac r a \right)+\frac{1-e^{2}}{n^{2}a e} W_{r} \cos f \right] \label{eq:Pbdfdt1}
\eeq
We aim at performing a first order calculation in the perturbation, which is supposed to be small (see e.g. \cite{brumberg,will13,iorioD}). As a consequence, first we may write
\beq
\dert t f \simeq  \left (\frac r a \right)^{2}\frac{1}{n\sqrt{1-e^{2}}}  \left[1+\frac{2}{n^{2}a} W_{r} \left(\frac r a \right)-\frac{1-e^{2}}{n^{2}a e} W_{r} \cos f \right] \label{eq:Pbdtdf1}
\eeq
and then, by integrating over one revolution along the unperturbed orbit (\ref{eq:ellipse1})  we obtain
\beq
P \simeq P_{b}+P_{A} \label{eq:Ppert1}
\eeq
where 
\beq
P_{b}\doteq \int_{0}^{2\pi}  \left (\frac r a \right)^{2}\frac{df}{n\sqrt{1-e^{2}}}=\frac{2\pi}{n} 
\label{eq:Pb1}
\eeq
is the unperturbed period, and
\beq
P_{A}\doteq \int_{0}^{2\pi}  \left (\frac r a \right)^{2}\left[\frac{2}{n^{2}a} W_{r} \left(\frac r a \right)-\frac{1-e^{2}}{n^{2}a e} W_{r} \cos f \right]\frac{df}{n\sqrt{1-e^{2}}} \label{eq:PA1}
\eeq
is the variation of the orbital period due to the perturbation.

The above equations (\ref{gaus_o}), (\ref{gaus_M}) and (\ref{eq:PA1}) allow to calculate the variations of the argument of pericentre,  mean anomaly and orbital period. This can be accomplished at least by numerical methods for arbitrary perturbations, however, as we are going to show in next Section,  when the perturbation is in the form of a power law,  it is possible to obtain analytic expressions.

\section{Power laws perturbations} \label{sec:powerlaw}

After having discussed in the previous Section the expressions for the secular perturbations of the orbital elements for a generic purely radial perturbation, here we focus on the particular case of perturbations in the form of a power law.  As we are going to show, in this case we can obtain analytic expressions for the secular variations. We point out that an arbitrary spherical symmetric perturbation that is expressed by an analytic function can be expanded in power series up to the required approximation level: hence, by knowing the contribution of each term of the series, it is possible to evaluate the whole effect of the perturbation, within the required accuracy. Eventually, even though our approach is motivated by the study of the effects of alternative theories of gravity, the results that we are going to obtain are quite general, and apply to radial perturbations of Keplerian orbits by means of arbitrary  power laws.

In particular, we consider two kinds of perturbation that we write as follows. Given a constant $\alpha$, which is a parameter deriving from the gravity model alternative to GR, the perturbations that we focus on are in the form:  (i) $\phi_{A}(r)=\alpha r^{N} $ where the integer $N$ is such that $N\leq -1$,  or differently speaking $\phi_{A}(r)=\frac{\alpha}{r^{|N|}}, \quad |N| \geq 1$; (ii) $\phi_{A}(r)=\alpha r^{N}$ where the integer $N$ is such that $N \geq 1$.
The perturbations of the first kind are asymptotically flat, while the second ones are not: in the latter case, we  assume that there exists a range $0<r<\bar r$ for which  $\phi_{A}(r) \ll \phi_{GR}(r)$,   and our results are valid within this range. We notice that, for a logarithmic perturbation in the form $\phi_{A}(r)=\beta \log (r/r_{0})$, it is $W_{r}=-\frac 1 2 \frac{\beta}{r}$, which can be dealt with as in the case (i) above.

In what follows we perform first-order calculations in the perturbation to obtain the secular variations; the latter can be used to place constraints on $\alpha$  by means of a comparison with the available data.

\subsection{Perturbation of the argument of pericentre} \label{ssec:omega}

For perturbations in the form $\phi_{A}(r)=\frac{\alpha}{r^{|N|}}, \quad |N| \geq 1$, we start from eq. (\ref{gaus_o}), and we average it over one orbital period by taking into account the expression of the unperturbed orbit (\ref{eq:ellipse1}) and the relation \cite{roy}
\beq dt = \frac{(1-e^2)^{3/2}}{n(1+e\cos f)^2} df \label{eq:dtdf00} \eeq
On using (\ref{eq:intc}), we then obtain
\beq
<\dot{\omega}>=-\frac 1 2 \frac{\pi \alpha |N|(|N|-1)\left(1-e^{2}\right)^{1-|N|}}{n^{2}a^{2+|N|}}  F\left(1-\frac{|N|}{2}, \frac 3 2 - \frac{|N|}{2}, 2,e^{2}\right)   \label{eq:secomega1}
\eeq
where $F$ is the hypergeometric function (see Appendix \ref{sec:integrals}).  In particular, this result is in agreement with the correspondent expression found in \cite{adkins}, where the effects of a central force were considered.

On the other hand, for perturbations in the form $\phi_{A}(r)=\alpha r^{N}$ with $N \geq 1$, we start again from  (\ref{gaus_o}) but, in order to average it over one orbital, it is useful to introduce the expression of the unperturbed orbit 
\beq
r=a(1-e\cos E) \label{eq:ellipse2}
\eeq
in terms of the eccentric anomaly $E$, which is is related to the
true anomaly $f$ by the following relations:
\beq \cos f = \rp{\cos E - e}{1-e\cos E},\quad \quad \sin f = \rp{\sqrt{1-e^2}\sin E}{1-e\cos E} \label{eq:csphiE}.\eeq
On using the relation \cite{roy}
\beq
dt=\frac 1 n \left( 1-e\cos E \right)dE \label{eq:dtdE00}
\eeq
and the integrals (\ref{eq:intc}) and (\ref{eq:int}), we obtain
\beq
<\dot{\omega}>=-\frac 1 2 \frac{\pi \alpha N\sqrt{1-e^{2}}}{n^{2}a^{2-N}}  \left[\left(N-1 \right) F\left(1-\frac{N}{2}, \frac{3}{2} - \frac{N}{2}, 2,e^{2}\right)+2   F\left(1-\frac{N}{2}, \frac 1 2 - \frac{N}{2}, 1,e^{2}\right) \right] \label{eq:secomega2zero}
\eeq
or, equivalently
\beq
<\dot{\omega}>=-\frac 1 2 \frac{\pi \alpha N\left(N+1\right)\sqrt{1-e^{2}}}{n^{2}a^{2-N}} F\left(1-\frac{N}{2}, \frac{1}{2} - \frac{N}{2}, 2,e^{2}\right)  \label{eq:secomega2}
\eeq
In particular, eq. (\ref{eq:secomega2}) is in agreement with the correspondent expression found in \cite{adkins}.

These general expressions can be used to reproduce known results.  For instance, on setting $N=1$ in eqs (\ref{eq:secomega2}) we obtain the variation of pericentre due to a constant perturbation acceleration
\beq
<\dot{\omega}>=-\frac{\pi \alpha }{n^{2}a}\sqrt{1-e^{2}} \label{eq:cost1}
\eeq
This result is in agreement with Sanders \cite{Sanders} (also reported by \cite{adkins}), who, working in MOND framework, considered the case of the constant acceleration to put constraints on the effects of anomalous acceleration by analyzing Mercury's advance of perihelion. 

The perturbation in the form $\phi_{A}(r)=\alpha r^{2}$ coincides with the effect of a cosmological constant $\Lambda$ \cite{kerr03},\cite{iorio08} in GR, while in the case of $f(R)$ gravity, it describes the vacuum solution in Palatini formalism \cite{ruggiero05,ruggierojcap}. In this case, by setting $N=2$ in eq. (\ref{eq:secomega2}) we get
\beq
<\dot{\omega}>=-\frac{3\pi\alpha}{n^{2}}\sqrt{1-e^{2}}\label{eq:lambda1}
\eeq
On setting $\alpha=-\frac 1 3 \Lambda$ and using the relation $n^{2}a^{3}=M$ which holds for the unperturbed orbit, eq. (\ref{eq:lambda1}) becomes
\beq
<\dot{\omega}>=\frac{\pi\Lambda a^{3}}{M}\sqrt{1-e^{2}} \label{eq:lambda11}
\eeq
in agreement with \cite{kerr03,ruggierojcap,iorioE,arakida}.

On setting $|N|=4$ in (\ref{eq:secomega1}), we  obtain the perturbation determined by a vacuum solution of  Ho\v{r}ava-Lifshitz gravity \cite{iorio09a,iorio09b}; by approximating the hypergeometric function, the variation of the argument of pericentre becomes
\beq
<\dot{\omega}>\simeq -\frac{6\pi\alpha}{n^{2}a^{6}}\frac{\left(1+\frac 1 4 e^{2} \right)}{\left(1-e^{2} \right)^{3}} \label{eq:omegaN4}
\eeq
in agreement with \cite{iorio09a}.

If we  consider a logarithmic perturbation, in the form $\phi_{A}(r)=\beta \log(r/r_{0})$, from eq. (\ref{eq:secomega1}), by approximation up to $e^{4}$ we obtain
\beq
<\dot{\omega}> =-\frac 1 2  \frac{\pi\beta\left(1-e^{2} \right)}{n^{2}a^{2}}F\left(1,\frac 3 2, 2,e^{2}\right)\simeq
 -\frac 1 2  \frac{\pi\beta\left(1-e^{2} \right)}{n^{2}a^{2}} \left(1-\frac 1 4 e^{2} \right)\label{eq:log1}
\eeq
in agreement with \cite{adkins}.

Eventually,  a perturbation in the form $\phi_{A}(r)=\frac{\alpha}{r^{2}}$ is obtained both in the case of Reissner-Nordstr\"om spacetimes \cite{iorio2012} and in recent works pertaining to the constraints for $f(T)$ gravity deriving from the Solar System observations \cite{ioriosari2012}.  On setting $|N|=2$ in (\ref{eq:secomega1}), we do obtain
\beq
<\dot{\omega}>=-\frac{\pi \alpha}{n^{2}a^{4}}\frac{1}{\left(1-e^{2} \right)} \label{eq:rnfT}
\eeq
in agreement with \cite{iorio2012,ioriosari2012}.

\subsection{Perturbation of the mean anomaly} \label{ssec:mean}

We may proceed as in the previous section to calculate the secular variation of the mean anomaly. However, we point out it is hard to use the mean anomaly in observational tests, because of the unavoidable uncertainty arising from the Keplerian mean motion $n$.

For a perturbation in the form $\phi_{A}(r)=\frac{\alpha}{r^{|N|}}, \quad |N| \geq 1$, we start from eq. (\ref{gaus_M}), and focus on the part that is proportional to $W_{r}$;  after averaging over the unperturbed ellipse (\ref{eq:ellipse1}) making use of eqs. (\ref{eq:dtdf00}) and (\ref{eq:int}), we get
\beq
<\dot{\mathcal{M}}>=-\frac{2 \pi\alpha |N| \left( 1-e^{2}\right)^{3/2-|N|}}{n^{2}a^{2+|N|}}   F\left(\frac 1 2-\frac{|N|}{2}, -\frac{|N|}{2},1,e^{2}\right)-\sqrt{1-e^{2}} <\dot{\omega}> \label{eq:secM1}
\eeq
where $<\dot{\omega}>$ is given by (\ref{eq:secomega1}). 

Similarly, for perturbations in the form  $\phi_{A}(r)=\alpha r^{N}$, with $N \geq 1$, averaging the part of (\ref{gaus_M}) which depens on $W_{r}$, making use of (\ref{eq:dtdE00}) and (\ref{eq:int}), we obtain 
\beq
<\dot{\mathcal{M}}>=\frac{2 \pi\alpha N}{n^{2}a^{2-N}}  F\left(-\frac{N}{2},-\frac 1 2  -\frac{N}{2},1,e^{2}\right)-\sqrt{1-e^{2}} <\dot{\omega}> \label{eq:secM2}
\eeq
where $<\dot{\omega}>$ is given by (\ref{eq:secomega2}).\\ 

On setting $N=2$, from eq. (\ref{eq:secM2}), by approximating the hypergeometric function, the variation of the mean anomaly is
\beq
<\dot{\mathcal{M}}>=\frac{4\pi \alpha}{n^{2}}F\left(-1,-\frac 3 2,1, e^{2}\right)+ \frac{3\pi\alpha}{n^{2}}\left(1-e^{2}\right)\simeq\frac{3\pi \alpha }{n^{2}} \left(\frac 7 3 +e^{2} \right)
\eeq
in agreement with \cite{ruggierojcap}.

\subsection{Perturbation of the orbital period} \label{ssec:period}

Starting from eq. (\ref{eq:PA1}), we may calculate the variation of the orbital period. In particular, for a perturbation in the form $\phi_{A}(r)=\frac{\alpha}{r^{|N|}}, \quad |N| \geq 1$, on 
evaluating eq. (\ref{eq:PA1}) over the unperturbed ellipse (\ref{eq:ellipse1}) making use of the integrals (\ref{eq:intc}), (\ref{eq:int}), we obtain

\beq
P_{A}= \frac{	\pi\alpha|N| \left(1-e^{2} \right)^{3/2-|N|}}{n^{3}a^{2+|N|}}\left[2 F\left(1-\frac{|N|} {2}, \frac 3 2-\frac{|N|}{2}, 1, e^{2} \right)-\frac 1 2 \left(|N|-1\right) F\left(1-\frac{|N|}{2}, \frac 3 2-\frac{|N|}{2}, 2, e^{2} \right) \right] \label{eq:secperiod1}
\eeq

As for perturbations in the form $\phi_{A}(r)=\alpha r^{N}$ with $N \geq 1$, on 
evaluating eq. (\ref{eq:PA1}) over the unperturbed ellipse (\ref{eq:ellipse2}) making use of
the relation (\ref{eq:dfdM1}) and the integrals (\ref{eq:intc}), (\ref{eq:int}), we obtain

\beq
P_{A}=-\frac{\pi\alpha N }{n^{3}a^{2-N}}\left[2F\left(-\frac N 2,-\frac 1 2 -\frac N 2,1,e^{2} \right)+\frac{\left(1-e^{2} \right)}{2} \left(N+1 \right)F\left(1-\frac N 2, \frac 1 2 -\frac N 2,2,e^{2} \right) \right] \label{eq:secperiod2}
\eeq

On setting $|N|=4$ , from eq. (\ref{eq:secperiod1}), by approximating the hypergeometric functions, we obtain
\beq
P_{A} \simeq  \frac{\pi\alpha  \left(1-e^{2} \right)^{-5/2}}{n^{3}a^{6}}\left(2+\frac 5 2 e^{2}  \right) \label{eq:periodN4}
\eeq
in agreement with  \cite{iorio09b}.
 
Furthermore, on setting $N=2$ in (\ref{eq:secperiod2}), by approximation of the hypergeometric functions, the perturbation of the orbital period turns out to be
\beq
P_{A}\simeq-\frac{\pi \alpha}{n^{3}}\left(7+3e^{2} \right) \label{eq:periodN2}
\eeq
in agreement with  \cite{iorio08}.

\section{Conclusions} \label{sec:conc}

We considered a general stationary and spherically symmetric  spacetime, and worked out the perturbations determined by a generic alternative theory of gravity to the GR solution describing the gravitational field around a central mass. In order to evaluate the effects of such perturbations, we showed that, in the weak-field and slow-motion limit,  to lowest approximation order, these effects can be described by a purely radial  acceleration. Then,  we  considered the Keplerian orbit of a test particle, which  can be thought of as a model of the dynamics of celestial bodies in planetary systems, and we evaluated the impact of such  perturbations on the orbital elements.

In particular, we obtained analytic expressions, in terms of hypergeometric functions, for the secular variations of the advance of pericentre, mean anomaly and orbital period determined by perturbations in form of power laws; the other orbital elements do not undergo secular variations to lowest approximation order. The expressions for the variation of the argument of pericentre are in agreement with the results recently obtained, pertaining to the orbital precession due to central force perturbations.  

Our results are quite general, since even though  were motivated by the study of the effect of alternative theories of gravity,  they can be  applied to arbitrary radial power laws perturbations of the orbital elements of a Keplerian motion.

Spherically symmetric perturbations in the form of a power law were obtained both in GR, for instance when the effect of a cosmological constant is considered, or in Reissner-Nordstr\"om spacetimes, and in alternative theories of gravity; we have shown that our results are in agreement with those already available in the literature. For an arbitrary perturbation it is possible to obtain the secular variations by means of numerical approaches, or by expanding it in power series and applying our results to each term of the series. 

The possibility of testing the predictions of alternative theories of gravity on planetary motion is important to verify their reliability on scales different from those typical of galactic dynamics or cosmology, where they are usually tested. 
The simple approach that we considered here allows to place bounds on the theory parameters, by evaluating  their impact on the Keplerian dynamics and allowing a comparison with the available data. In this context, it is very important the development in the study of Solar System ephemerides, due to the works of Russian \cite{pit1,pit2} and French \cite{fienga} research teams.

\appendix
\section{Solutions of Integrals by means of Hypergeometric Functions} \label{sec:integrals}

In order to evaluate the secular variations of the Keplerian elements and of the orbital period determined by the perturbations in the form of power laws that we have considered, it is necessary to solve the following integrals

\beq 
I_{N}=\int_{0}^{2\pi} \cos u \left[1+e\cos u \right]^{N}du  \label{eq:intc0}
\eeq

\beq 
L_{N}=\int_{0}^{2\pi} \left[1-e\cos u \right]^{N}du \label{eq:int0}
\eeq
They can be solved by subsituting $\cos u= z$, and then using the binomial theorem for $\left(1+z\right)^{N}$. The solutions are given in terms of hypergeometric functions:
\beq 
I_{N}=\pi N e F\left(\frac 1 2- \frac N 2 , 1 - \frac N 2 , 2, e^{2} \right)   \label{eq:intc}
\eeq
\beq 
L_{N}=2 \pi  F\left(- \frac N 2 , \frac 1 2 - \frac N 2 , 1, e^{2} \right)   \label{eq:int}
\eeq
where  $F(a,b,c,x) \doteq _{2}F_{1}(a,b,c,d)$ defined by (see e.g. \cite{arf})
\beq
_{2}F_{1}(a,b,c,x)=\sum_{n=0}^{\infty} \frac{(a)_{n}(b)_{n}}{(c)_{n}}\frac{x^{n}}{n!} \label{eq:hyp1}
\eeq
where the Pochhammer  symbol $(a)_{n}$ is defined by
\begin{eqnarray}
(a)_{n} & \doteq & \frac{\left(a+n-1\right)!}{\left(a-1 \right)!} \label{eq:hyp2a} \\
(a)_{0} & \doteq & 1 \label{eq:hyp2b}
\end{eqnarray}

\end{document}